\documentclass[review,authoryear]{elsarticle}

\usepackage{hyperref}

\journal{Information Processing \& Management}

\makeatletter
\renewcommand\@biblabel[1]{#1.}
\makeatother

\usepackage{paralist}
\usepackage{bm}
\usepackage{footnote}
\usepackage{amsfonts}
\usepackage{multirow}
\usepackage{enumerate}
\usepackage{amssymb}
\usepackage[skip=0pt]{subcaption}
\usepackage{graphicx}
\usepackage[skip=0pt]{caption}
\usepackage{booktabs}
\usepackage{color}
\usepackage{colortbl}
\usepackage{url}
\usepackage{mathtools}
\usepackage{setspace}
\usepackage{algorithm}
\usepackage{algorithmic}
\usepackage{enumitem}

\newcommand{\noop}[1]{}

\newcommand{\todo}[1]{\textcolor{black}{#1}}

\allowdisplaybreaks

\newcommand{\argmax}{\operatornamewithlimits{arg\,max}}

\usepackage{acronym}
\acrodef{RNN}{recurrent neural network}
\acrodef{CNN}{convolutional neural network}
\acrodef{LSTM}{long and short-term memory network}
\acrodef{Tree-LSTM}{tree-structured long short-term memory networks}
\acrodef{LIR}{Local Interaction Representation}
\acrodef{GIR}{Global Interaction Representation}
\acrodef{HIR}{Hybrid Interaction Representation}
\acrodef{HLMPf}{hybrid language model pretrain-finetuning}
\acrodef{PIF}{Pre-train, Interact, Fine-tune}
\acrodef{RER}{Relative Error Rate}

\biboptions{authoryear}

\begin{document}

\begin{frontmatter}

\title{\acl{PIF}: A Novel Interaction Representation for Text Classification}


\author[mymainaddress]{Jianming Zheng}\ead{ADDRESS}
\author[mymainaddress]{Fei Cai\corref{mycorrespondingauthor}}
\cortext[mycorrespondingauthor]{Corresponding Author}\ead{caifei@nudt.edu.cn}
\author[mymainaddress]{Honghui Chen}\ead{ADDRESS}
\author[mysecondaryaddress]{Maarten de Rijke}\ead{derijke@uva.nl}
\address[mymainaddress]{Science and Technology on Information Systems Engineering Laboratory, \\National University of Defense Technology, Hunan, 410073, China}
\address[mysecondaryaddress]{Informatics Institute, University of Amsterdam, Amsterdam, 1098 XH, The Netherlands}



\begin{abstract}
Text representation can aid machines {in} understanding text.
Previous work on text representation often focuses on the so-called forward implication, i.e., preceding words are taken as the context of later words for creating representations, thus ignoring the fact that the semantics of a text segment is a product of the mutual implication of words in the text: later words contribute to the meaning of preceding words. 
We introduce the concept of interaction and propose a two-perspective interaction representation, that encapsulates a local and a global interaction representation.
Here, \todo{a \emph{local} interaction representation is one that interacts among words with parent-children relationships on the syntactic trees and a \emph{global} interaction interpretation is one that interacts among all the words in a sentence.}
We combine the two interaction representations to develop a \acf{HIR}.

Inspired by existing feature-based and fine-tuning-based pretrain-finetuning approaches \todo{to language models }, we integrate the advantages \todo{of feature-based and fine-tuning-based methods} to propose the \acf{PIF} architecture.

We evaluate our proposed models on five widely-used datasets for text classification tasks. 
Our ensemble method, HIR$_P$, outperforms state-of-the-art baselines with improvements ranging from  2.03\% to 3.15\% in terms of \todo{error rate}.
In addition, we find that, \todo{the improvements of \ac{PIF} against most state-of-the-art methods} is not affected by increasing of the length of the text.
\end{abstract}

\begin{keyword}
Interaction representation, Pre-training, Fine-tuning, Classification
\end{keyword}

\end{frontmatter}


\section{Introduction}
Text representations map text spans into real-valued vectors or matrices. 
They have come to play a crucial role in machine understanding of text. 
Applications include sentiment classification~\citep{Tang2015Document}, question answering~\citep{Qin2017Enhancing}, summarization~\citep{Ren2017Leveraging}, and sentence inference~\citep{Parikh2016A}.

Previous work on text representation can be categorized into three main types ~\citep{Xie2016I}, i.e., \emph{statistics-based}, 
\emph{neural-network-based} and \emph{pre\-training-based} embeddings. 
Statistics-based embedding models are estimated based on a statistical indicator, e.g., the frequency of co-occurring words (in bag-of-words models~\citep{Joachims1998Text}), the frequency of co-occur\-ring word pairs (in n-gram models~\citep{Zhang2015Character}), and the weights of words in different documents (the TF-IDF model~\citep{Robertson2004Understanding}).
{Neural-network-based embedding models} mainly rely on a neural network architecture to learn a text representation, based on a hidden layer~\citep{Joulin2016Bag}, \acp{CNN}~\citep{Kim2014Convolutional} or \acp{RNN}~\citep{Liu2016Recurrent}.
Additionally, this type of methods may also consider the syntactic structure to reflect the semantics of text, e.g., recursive neural networks~\citep{Socher2013EMNLP} and \ac{Tree-LSTM}~\citep{Tai2015improved}. 
Pretraining-based embedding models adopt a feature-based~\citep{Mikolov2013Efficient,Pennington2014G,McCann2017L,Peters2018D} or fine-tuning strategy~\citep{Dai2015S,Howard2018U,Devlin2019B,Yang2019X} to capture the semantics and syntactic information from a large text corpora.

In general, the aforementioned models work well for the task of text classification. 
\todo{\citep{Joulin2016Bag,Kim2014Convolutional,Zhang2015Character,Howard2018U}}
However, in existing embedding models, the \todo{ generated process of the vectorized representation of a text} usually follows a so-called one-way action.
That is to say, representations generated for the preceding text are taken as the context to determine the representations of later texts.
Although a bidirectional LSTM considers bidirectional actions, it simply concatenates two one-way actions to get the embeddings. 
We argue that the semantics as defined in terms of a text representation should be a product of interactions of all source elements (e.g., words or sentences) in the text.
Restrictions to one-way actions may result in a partial semantic loss~\citep{Saif2016C}, \todo{causing the poor performances in the downstream applications.}
We hypothesize that although these interaction relations may be learned by neural networks with enough samples, explicitly modeling such interaction relations can directly make text representation more informative and effective.
Furthermore, recent unsupervised representation learning has proven to be effective and promising in the field of natural language processing~\citep{McCann2017L, Peters2018D, Howard2018U,Devlin2019B,Yang2019X}.
So far, these approaches are limited to a single strategy (either feature-based or fine-tuning strategy), which results in a so-called fine-tune error, \todo{which may be trapped in the local best.}
\begin{figure}[t]
  \centering
   \includegraphics[width=0.6\columnwidth]{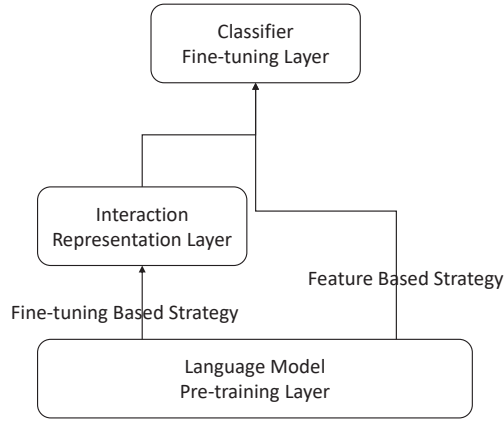}
   \caption{Overview pipeline of Pre-train Interact Fine-tune}
\label{overview pipeline}
\end{figure}

Thus, as illustrated in Figure~\ref{overview pipeline}, we focus on the task of text classification and propose a novel pipeline with the following ingredients:
\begin{enumerate}
   \item \textbf{pre-train} language model on a large text corpus to get the related word embeddings and neural networks parameters;
   \item \textbf{interact} the word embeddings based on the pre-trained parameters to obtain the interaction representation; and
   \item \textbf{fine-tune} the classifier with the interaction representation and pre-trained word embeddings as input.
\end{enumerate}

\noindent%
More specifically, in the interaction representation layer, we propose a two-perspective interaction representation using a \acfi{LIR}\acused{LIR} and a \acfi{GIR}\acused{GIR}.
The \ac{LIR} applies an attention mechanism \citep{Bahdanau2015Neural} inside the syntactic structure of a sentence, e.g., the dependency-based parse trees or constituency-based parse trees, to reflect the local interaction of adjacent words.
The \ac{GIR} employs an attention mechanism with an enumeration-based strategy to represent the interactions of all words in a sentence. 
After that, we combine \ac{LIR} and \ac{GIR} to into a \acfi{HIR}\acused{HIR} model to represent both local and global interactions of words in a sentence. 
For the pretrain-finetuning process, we combine the feature-based and the fine-tuning strategies and propose a \acfi{HLMPf}\acused{HLMPf} approach. 
\ac{HLMPf} first follows the fine-tuning  strategy to employ the pre-trained embeddings and neural network parameters as the initialization of the interaction representation layer.
Then, according to the feature-based strategy, \ac{HLMPf} applies the pre-trained embeddings as additional features and concatenates the interaction representation in the classifier fine-tuning layer.

For evaluation, we conduct a comprehensive experiment on five publicly available benchmark datasets for the task of text classification.
The experimental results show that our proposal with interaction representations and the hybrid pretrain-finetuning strategy outperforms the state-of-the-art baselines for text classification, with improvements ranging from 2.03\% to 3.15\% in terms of accuracy. 

The main contributions of our work are as follows: 
\begin{enumerate}
\item We propose a novel pipeline for the task of text classification, i.e., \acfi{PIF}.
\item To the best of our knowledge, ours is the first attempt to model word interactions for text representation.
We introduce a two-perspective interaction representation for text classification, i.e., a \acf{LIR} and a \acf{GIR}, which are then combined to generate a \acf{HIR} model.
\item We combine the advantages of two popular language model pretrain-fine-tuning strategies (feature-based and fine-tuning) and propose the \acf{HLMPf}.
\item We analyze the effectiveness of our proposal and find that it outperforms the state-of-the-art methods for text classification in terms of accuracy. 
\end{enumerate}

\section{Related Work}
\label{Related Work}
In this section, we briefly summarize the general statistical approaches for text representation in Section~\ref{Statistical} and the neural-networks-based methods in Section~\ref{Neural}.
We then describe the recent work on language model pre-training for downstream applications in Section~\ref{language}.

\subsection{Statistics-based representation}
\label{Statistical}
As a word is the most basic unit of semantics, the traditional one-hot representation model converts a word in a vocabulary into a sparse vector with a single high value (i.e., 1) in its position and the others with a
low value (i.e., 0). 
The representation is employed in the Bag-of-Words (BoW) model~\citep{Joachims1998Text} to reflect the word frequency.
However, the BoW model only symbolizes the word and cannot reflect the semantic relationship between words.
Consequently, the bag-of-means model~\citep{Zhang2015Character} was proposed to cluster the word embeddings learned by the word2vec model~\citep{Mikolov2013Efficient}. 
Furthermore, the bag-of-n-grams~\citep{Zhang2015Character} was developed to take the n-grams (up to 5-grams) as the vocabulary in the BoW model.
In addition, with some extra statistical information, e.g., TF-IDF, a better document representation can be produced~\citep{Robertson2004Understanding}.
Other text features, e.g., the noun phrases~\citep{Lewis1992An} and the tree kernels~\citep{Post2013Explicit}, were incorporated into the model construction.

Clearly, a progressive step has been made in statistical based representation~\citep{Bernauer2018T}.
However, such traditional statistical representation approaches inevitably face the problems of data sparsity and dimensionality, leading to no applications on large-scale corpora.
In addition, such approaches are simply built on shallow statistics, and a deeper semantic information of the text has not been well developed. 

Instead, our proposal in this paper based on neural networks has the ability to learn a
low-dimensional and distributed representation to overcome such problems.

\subsection{Neural-based representation}
\label{Neural}
Since \citet{Bengio2013A} first employed the neural network architecture to train a language model, considerable attention has been devoted to proposing neural network-related models for text representation.
For instance, the FastText model~\citep{Joulin2016Bag} employs one hidden layer to integrate the subword information and obtains satisfactory results.
However, this model simply averages all word embeddings and discards the word order. 
In view of that, \citet{Liu2016Recurrent} employed the recurrent structure, i.e., \acp{RNN}, to consider the word order and to jointly learn {text representation} across multiple related tasks. 
Compared to \acp{RNN}, \acp{CNN} are easier to train and capture the local word-pair information~\citep{Kim2014Convolutional,Zhang2015Character}.

Furthermore, a combination of neural network models are integrated to develop the advantage of each single neural network.
For example, \citet{Lai2015Recurrent} proposed the recurrent convolutional neural networks (RCNN), which adopted the recurrent structure to grasp the context information and employed a max-pooling layer to identify the key components in text.
Besides, other document features have been injected into the document modeling.
For instance, \citet{Zheng2019C} took the hierarchical structure of text into account. \citet{He2018E} transformed the document-level knowledge to improve the performance of aspect-level sentiment classification.

Although these approaches have been proved effective in the downstream applications, they completely depend  on the structure of network to implicitly represent a document, ignoring the  interaction that exists among the source elements in a document, e.g., words or sentences.
However, our proposal can model the interaction as the starting point to better reflect the semantic relationship between words in a sentence, which we argue can help improve the performance of downstream tasks, e.g, sentimental classification. 

\subsection{Language model pre-training-based representation}
\label{language}
The language pre-training model has been shown effective for the natural language processing tasks, e.g., question answering~\citep{McCann2017L}, textual entailment \citep{Peters2018D}, semantic role labeling \citep{Devlin2019B} sentimental analysis \citep{Dai2015S}, etc.
These pre-training models can be mainly classified into two classes, i.e., feature-based models and fine-tuning models.

The feature-based models generate the pre-trained embeddings from other tasks, where the output can be regarded as the additional features for the current task architecture.
For instance, word2vec \citep{Mikolov2013Efficient} and GloVe \citep{Pennington2014G} focus on transforming words into the distributed representations and capturing the syntactics as well as the semantics by pre-training the neural language models on a large text corpora. 
In addition, \citet{McCann2017L} concentrated on the machine translation task to get the contextualize word vectors (CoVe).
Since these word-level models suffer from the word-ploysemy, \citet{Peters2018D} developed the sequence-level model, i.e., ELMo, to capture the complex word features across different linguistic contexts and then use ELMo to generate the context-sensitive word embeddings.

Different from the feature-based strategy~\citep{Mehta2018E}, the fine-tuning models first produce the contextual word presentations which have been pre-trained from unlabeled text and fine-tune for a supervised downstream task. For instance,
\citet{Dai2015S} trained a sequence auto-encoder model on unlabeled text as an initialization of another supervised network.
However, this method suffers from overfitting and requires some in-domain knowledge to improve the performance.
Consequently, Universal Language Model Fine-tuning (ULMFit) \citep{Howard2018U} was developed, which leveraged the general-domain pre-training and the novel fine-tuning techniques to prevent overfitting.
In addition, \citet{Devlin2019B} proposed {two unsupervised tasks, i.e., masked language model and next sentence prediction, to further improve fine-tuning process.}
In addition, XLNet \citep{Yang2019X} was proposed to employ the permutation language model to capture the bidirectional context and avoid the pretrain-finetune discrepancy.

Although the language pre-training model based representations have been proposed and proved promising in the NLP tasks, these methods are limited to either feature-based or fine-tune-based strategy.
Our proposal combine their respective characteristics to improve the performance of downstream applications.

\section{Proposed Models}
In this section, we first formally describe how to compute the interaction representation in Section~\ref{IR_D}, which can be divided into three parts, i.e., \ac{LIR} (see Section~\ref{LIR_D}),\ac{GIR} (see Section \ref{GIR_D}) and \ac{HIR} (see Section~\ref{HIR_D}).
And then, we introduce the \ac{HLMPf} approach in detail (see Section~\ref{hlmpf}), which is the combination of the feature-based and fine-tuning strategies.

\subsection{Interaction representation}
\label{IR_D}
We describe the \acf{LIR} of adjacent words and introduce the \acf{GIR} of all words in a sentence. After that, a \acf{HIR} model is proposed. 

\subsubsection{Local interaction representation}
\label{LIR_D}

We introduce an attentive tree LSTM that computes a local representation of words. The idea of an \emph{action} of a word on another word is that the former assigns a semantic weight to the latter.
 
\begin{figure}[t]
  \centering
   \includegraphics[width=0.95\columnwidth]{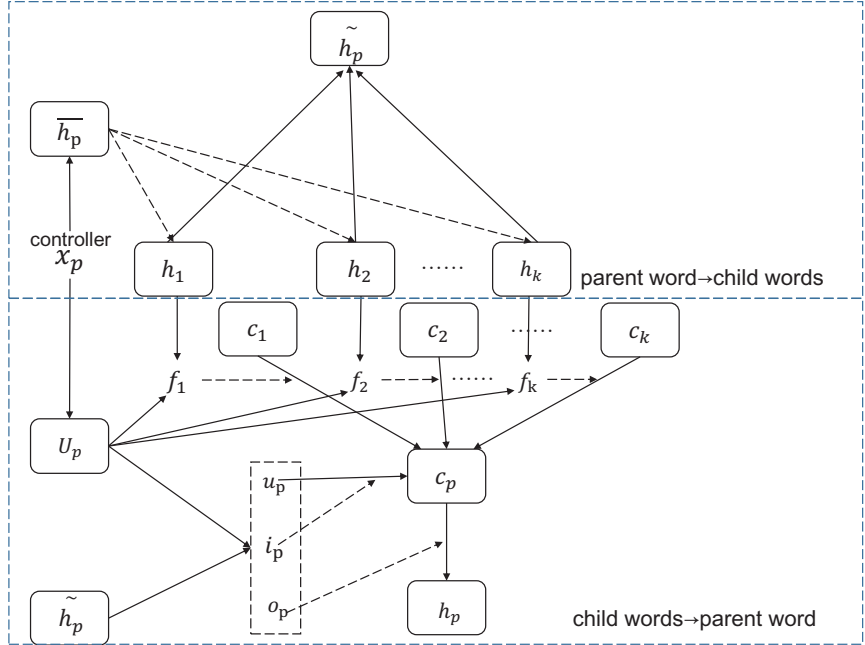}
   \caption{Structure of the local interaction representation model. (For simplicity, we write $U_p$ for $(U^{(i)}P_{x_p}$, $U^{(o)}P_{x_p}$, $U^{(u)}P_{x_p}$, $U^{(f)}x_p)$.)}
\label{Attentive Tree LSTM}
\end{figure}

The experiments we conduct related to LIR are based on con\-stituency-based trees, but we explain the core concepts for both dependency-based and constituency-based trees.
%
Given a depen\-dency-based parse tree, let $C(p)$ denote the set of child words of a parent word $x_p$. 
To define the attentive tree LSTM, we introduce hidden states and memory cells $h_k$ and $c_k$ $(k\in\{1$, $2$, \ldots, $|C(p)|\})$ for every child word, respectively.
As shown in Fig.~\ref{Attentive Tree LSTM}, unlike the \ac{Tree-LSTM} model in \citep{Tai2015improved} that only performs the one-way action (child words $\mapsto$ parent word), \ac{LIR} also considers an action in the opposite direction, i.e., parent word $\mapsto$ child words.

Let us explain this in detail. In an action parent word $\mapsto$ child words, we regard the parent word $x_p$ as a controller that assigns semantic weights based on the attention mechanism to its child words in a sentence~\cite{Saraiva2016A}.
Thus, we first convert the parent word $x_p$ into a hidden representation $\overline{h_p}$ as follows:
\begin{equation}
\overline{h_p} = \tanh(W^{(h)} P_{x_p} + b^{(h)}),
\label{eq_1}
\end{equation}
where $P_{x_p}$ is the pre-trained word embedding for parent word $x_p$; $W^{(h)}$ and $b^{(h)}$ are the weight matrix and the bias term, respectively.
Then, we employ a general content-based function \citep{Luong2015Effective} to connect the parent word and the child words as follows:
\begin{equation}
\alpha_k =\overline{h_p} W_\alpha h_k,
\label{eq_2}
\end{equation}
where $\alpha_k$ is the connective representation of $\overline{h_p}$ and the hidden state $h_k$, and $W_\alpha$ is the connective matrix to be learned.
After that, we apply a softmax function on a sequence of connective representations $\{\alpha_1,\alpha_2,\ldots,\alpha_{|C(p)|}\}$ to get the weight $\lambda_k$ as follows:
\begin{equation}
\lambda_k= \frac{\exp(\alpha_k)}{\sum_{i=1}^{|C(p)|}\exp(\alpha_i)}.
\label{eq_3}
\end{equation}
{Finally}, we represent the hidden interaction state $\tilde{h}_p$ that relates to all child states of the parent word $x_p$, i.e.,
\begin{equation}
\tilde{h}_p=\sum_{i \in C(j)}\lambda_i h_i.
\label{eq_4}
\end{equation}
In the action child words $\mapsto$ parent word in Fig.~\ref{Attentive Tree LSTM}, we use the hidden interaction state $\tilde{h_p}$ and the parent word $x_p$ as  input to the LSTM cell and obtain
\begin{eqnarray}
i_p &=& \sigma (U^{(i)} x_p + W^{(i)} \tilde{h_p} + b^{(i)}),
\label{eq_5}
\\
o_p &=& \sigma (U^{(o)} x_p + W^{(o)} \tilde{h_p} + b^{(o)}),
\label{eq_6}
\\
u_p &=& \tanh (U^{(u)} x_p + W^{(u)} \tilde{h_p} + b^{(u)}),
\label{eq_7}
\\
f_{kp} &=& \sigma (U^{(f)} x_p + W^{(f)} h_k + b^{(f)}),
\label{eq_8}
\end{eqnarray}
where $i_p$, $o_p$ and $f_{kp}$ are the input gate, the output gate and the forget gate, respectively; $u_p$ is the candidate hidden state of $x_p$.
For $i_p$, $o_p$, $u_p$ and $f_{kp}$, we have a corresponding weight matrix of $x_p$ (i.e., $U^{(i)}$, $U^{(o)}$, $U^{(u)}$ and $U^{(f)}$), a weight matrix of $\tilde{h}_p$ (or~$h_k$) (i.e., $W^{(i)}$, $W^{(o)}$, $W^{(u)}$ and $W^{(f)}$), and a bias term (i.e., $b^{(i)}$, $b^{(o)}$, $b^{(u)}$ and $b^{(f)}$).
Finally, we can get the memory cell $c_p$ and the hidden state $h_p$ of the parent word $x_p$ as follows:
\begin{eqnarray}
c_p&=& i_p \odot u_p + \sum_{k=1}^{|C(p)|} f_{kp} \odot c_k,
\label{eq_9}
\\h_p &=& o_p \odot \tanh(c_p),
\label{eq_10}
\end{eqnarray}
where $\odot$ is element-wise multiplication and $c_k$ is the memory cell of a child word.

Similarly, given a constituency-based tree, let $x_l$ and $x_r$ denote the left child word and the right child word of a parent word $x_p$.
Since the parent word $x_p$ is a non-terminal node (i.e., $x_p$ is a zero vector), we use $x_l$ and $x_r$ as the controller instead of $P_{x_p}$, respectively.
Therefore, following Eq.~\eqref{eq_2}--\eqref{eq_4}, we obtain the hidden interaction states $\tilde{h}_l$ and $\tilde{h}_r$ related to $x_l$ and $x_r$, respectively.
We concatenate $\tilde{h}_l$ and $\tilde{h}_r$ to represent the hidden interaction states of the parent word, i.e., $\tilde{h}_p=[\tilde{h}_l;\tilde{h}_r]$.
Again, following Eq.~\eqref{eq_5}--\eqref{eq_10}, we can get the memory cell $c_p$ and the hidden state $h_p$ for parent word $x_p$.

At this stage we have represented the local interaction, and each word has been updated by the interaction representation.

\subsubsection{Global interaction representation}
\label{GIR_D}
Unlike \ac{LIR}, which captures the syntactic relation between words, \ac{GIR} adopts a enumeration-based strategy to employ an attention mechanism on all words in a sentence.

In detail, after implementing \ac{Tree-LSTM} on all $n$ words in a sentence, we can have the hidden representations $\{h_1, h_2, \ldots, h_n\}$ corresponding to the words $\{x_1, x_2, \ldots, x_n\}$.
In order to represent the interaction between a word $x_g$ and the other words in a sentence, we regard the word $x_g$ as a controller that can assign semantic weights to other words in $\{x_1, x_2, \ldots, x_n\}$ excluding $x_g$ itself. Similarly, we employ a general content-based function to connect the word $x_g$ with other words as follows:
\begin{equation}
\alpha_{gk} = h_g W_\alpha h_k,
\end{equation}
where $\alpha_{gk}$ is the connective representation of  $h_g$ and $h_k$ $(g,k \in (1,2,\ldots,n))$.
After that, we can get all connective representations $\{\alpha_{g1},\alpha_{g2},\ldots,\alpha_{gn}\}$ between the word $x_g$ and other words.
Then, we can apply a softmax function on the connective representation sequence to calculate the weight as follows:
\begin{equation}
\lambda_{gk}=\frac{\exp(\alpha_{gk})}{\sum_{i=1}^{n} \exp(\alpha_{gi})},
\end{equation}
where $\lambda_{gk}$ is the weight of word $x_k$ in $\{x_1, x_2, \ldots, x_n\}$ that interacts with word $x_g$.
Finally, we obtain the interaction representation $r_g$ as follows:
\begin{equation}
r_g = \sum_{i=1}^{n} \lambda_{gi} h_i.
\end{equation} 
By doing so, we enumerate all words in a sentence and can return a sequence of interaction representations as $\{r_1,r_2,\ldots,r_n\}$.
We then adopt a max-pooling on this sequence to produce the sentence embeddings $s$ by 
\begin{equation}
s=\max\{r_1,r_2,\ldots,r_n\}.
\end{equation}
This completes the definition of the global interaction representation. 
We can train the sentence representation $s$ to update the pre-trained embeddings.

\subsubsection{Hybrid interaction representation}
\label{HIR_D}

In order to capture both local and global interactions between words, we combine \ac{LIR} and \ac{GIR} to form a hybrid interaction representation model (\ac{HIR}) for text representation.
\ac{HIR} first follows the procedure of \ac{LIR} to produce the hidden state representations $\{h_1,h_2,\ldots,h_n\}$ for the corresponding word $\{x_1,x_2,\ldots,x_n\}$.
Then, \ac{HIR} employs the process of \ac{GIR} on these hidden state representations to get the final sentence embeddings $s$.

Eventually, in the process of class prediction, we apply a softmax classifier on the sentence embeddings $s$ to get a predicted label $\hat{s}$, where $\hat{s} \in \mathcal{Y}$ and $\mathcal{Y}$ is the class label set, i.e.,
\begin{equation}
\hat{s}=\argmax p(\mathcal{Y}\mid s),
\label{pred}
\end{equation}  
where 
\begin{equation}
p(\mathcal{Y}\mid s)=\mathrm{softmax}(W^{(s)} s + b^{(s)}).
\label{pred2}
\end{equation}
Here, $W^{(s)}$ and $b^{(s)}$ are the reshape matrix and the bias term, respectively.
For formulating the loss function in \ac{HIR}, we combine the corresponding loss in \ac{LIR} and \ac{GIR} as 
\begin{equation}
L=\frac{\gamma}{n} \sum_{i=1}^{n} \log~p(\tilde{w_i}\mid h_i)-(1-\gamma)\log~p(\tilde{s}\mid s),
\end{equation}
where the former loss comes from LIR and the latter from GIR, $\gamma$ is the trade-off parameter.
In addition, $h_i$ is the hidden state and $\tilde{w_i}$ is the true class label of word $x_i$ in LIR; $\tilde{s}$ is the true class label of sentence embeddings $s$ in GIR.  
In addition, $\tilde{w_i}$ and $\tilde{s}$ can be trained using the dataset.

We have now introduced the main process of our \ac{HIR} model. 
Clearly, as shown in Algorithm \ref{algorithm1}, we first employed bi-lstm process the pre-trained word sequence to build their semantics relations from step 1 to 2. 
Then, with the help of syntactic parse tool, we can get the parent-child set $T$. 
Following the bottom-up traversal algorithm, we show how to model the local interaction representation between parent word and child words from step 4 to 11.
While from step 12 to 19, we show how to compute the global interaction representation between all words.
At last, we optimize the loss function to jointly training the process of LIR and GIR.
And update the pre-trained embeddings with the interaction representations.

\begin{algorithm}[!t]
\caption{{Hybrid Interaction Representation}}
\label{algorithm1}
\begin{algorithmic}[1]
\REQUIRE The pre-trained embeddings for each word in a sentence $s_t$, i.e., $\{P_{x_1},P_{x_2}, \ldots, P_{x_n}\}$; the pre-trained parameters $\psi$ of the neural networks from the language pre-training layer.
\ENSURE The interaction representation for the word sequence, i.e., $\{I_{x_1},I_{x_2}, \ldots, I_{x_n}\}$
\STATE $\overrightarrow{h_t} \gets \overrightarrow{LSTM} (x_t, h_{t-1})$, $\overleftarrow{h_t} \gets \overleftarrow{LSTM} (x_t, h_{t-1})$
\STATE $h_t= (\overrightarrow{h_t};\overleftarrow{h_t}), t=1,2, \ldots,n$
\STATE Get the set of parent word $p_i$ and its child words by syntactic parsing: \\
$T=\{p_i, C(p_i)\} \gets $ syntactic parse ($s$), $i=1,\ldots,|T|$
\FOR { each parent word $\{p_i, C(p_i)\} \in T$}
\STATE $\overline{h_{p_i}} = \tanh(W^{(h)} P_{x_{p_i}} + b^{(h)})$
     \FOR {each child word in $C(p_i)$:} 
         \STATE Get the connective representation: $\alpha_k =\overline{h_p} W_\alpha h_k$
     \ENDFOR
     \STATE $\tilde{h}_p=\sum_{i \in C(j)}\lambda_i h_i$, where $\lambda_k= \frac{\exp(\alpha_k)}{\sum_{i=1}^{|C(p)|}\exp(\alpha_i)}$ 
\\ ~~~~~~~~~~~~~~~~~~~~~~~~~~~~~~~~~~~~~\%\% parent word $\mapsto$ child words 
\STATE $h_p \gets LSTM (x_p, \tilde{h_p})$ \%\% child words $\mapsto$ parent word
\ENDFOR \\\%\% \textit{This loop for LIR that follows the bottom-up algorithm to traverse the syntactic parsing tree.} 
\FOR {each word $w_g$ in sentence $s_t$}
\STATE Word $w_g$ is regarded as the parent word
\FOR {each word in sentence $s_t$ excluding word $w_g$}
\STATE $\alpha_{gk} = h_g W_\alpha h_k$
\ENDFOR
\STATE $r_g = \sum_{i=1}^{n} \lambda_{gi} h_i$, where $\lambda_{gk}=\frac{\exp(\alpha_{gk})}{\sum_{i=1}^{n} \exp(\alpha_{gi})}$
\STATE $s=\max\{r_1,r_2,\ldots,r_n\}$
\ENDFOR ~~~\%\% \textit{This loop for GIR.}
\STATE Optimize the loss function: $L=\frac{\gamma}{n} \sum_{i=1}^{n} \log~p(\tilde{w_i}|h_i)-(1-\gamma)\log~p(\tilde{s}|s)$
\STATE Update $I_{x_i} \gets h_{x_i}$
\RETURN $\{I_{x_1},I_{x_2}, \ldots, I_{x_n}\}$
\end{algorithmic}
\end{algorithm}

\subsection{Hybrid language model pretrain-finetuning}

\label{hlmpf}
Unsupervised representation learning, as a fundamental tool, has been shown effective in many language processing tasks~\citep{McCann2017L, Peters2018D, Howard2018U,Devlin2019B,Yang2019X}.
%
%
Here, we propose the \acfi{HLMPf} method, which integrates the respective advantages in the \ac{PIF} pipeline shown in Figure \ref{overview pipeline}. The details of HLMPf are shown in Algorithm \ref{algorithm 2}.

We first follow the BERT \citep{Devlin2019B} model to train the language model pre-training layer.
From step 2 to 3, we employ the fine-tuning strategy to fine-tune the interaction representation layer and the language model pre-training layer.
After that, we follow the ELMo approach \citep{Peters2018D} to obtain the context-aware word embeddings.
From step 5 to 6, we show how to further fine-tune all neural layers following the feature-based strategy.

Specially, since fine-tuning all layers at once will result in catastrophic forgetting, we adopt the gradual unfreezing strategy \citep{Howard2018U} to fine-tune all neural layers.

\begin{algorithm}[h]
\caption{Hybrid Language Model Pretrain-finetuning }
\label{algorithm 2}
\begin{algorithmic}[1]
\REQUIRE The text need to be trained.
\ENSURE The trained parameters $\psi$ of all neural networks; the trained word embeddings $\{W_{x_1},W_{x_2}, \ldots, W_{x_n}\}$ .
\STATE Pre-train the masked language model and next sentence prediction tasks to get the pre-trained neural networks and related parameters.
\STATE Add the interaction representation layer to the pre-training layer.
\STATE Following algorithm \ref{algorithm1}, optimize the loss function to fine-tune the related parameters. \\~~~~~~~~~~~~~~~~~~~~~~~~~~~~~~~~~~~~~~~~~~\%\%  \textbf{the fine-tuning strategy}
\STATE Pre-train some supervised tasks to get the context-aware word embeddings, i.e., $\{C_{x_1},C_{x_2}, \ldots, C_{x_n}\}$.
\STATE Add the classifier fine-tuning layer to the former combination layer.
\STATE Use the $\{I_{x_I}; C_{x_i}\}$ as the input of the classifier fine-tuning layer to further fine-tune the related parameters.
 \\~~~~~~~~~~~~~~~~~~~~~~~~~~~~~~~~~~~~~~~\%\%  \textbf{the feature-based strategy}
\RETURN $\psi$ and $\{W_{x_1},W_{x_2}, \ldots, W_{x_n}\}$
\end{algorithmic}
\end{algorithm}

\section{Experiments}
\label{exp}

We start by providing an overview of the text representation model to be discussed in this paper and list the research questions that guide our experiments.
Then we describe the task and datasets that we evaluate our proposals on. 
We conclude the section by specifyingthe settings of the parameters in our experiments.

\begin{table*}[!t]
  \centering

\begin{footnotesize}
\caption{An overview of models discussed in the paper.}
    \begin{tabular}{@{}lp{5cm}l@{}c@{}}
    \toprule
Model & Description & Source & Finetuning \\
 \midrule
LSTM & \raggedright A \acf{LSTM} based representation model. &\citep{Lai2015Recurrent} &$\times$\\
C-CNN & \raggedright A \ac{CNN} based representation model in the character level. & \citep{Zhang2015Character} &$\times$\\
CoVe & \raggedright A text representation model transferred from the machine translation model. & \citep{McCann2017L}&$\surd$\\
ULMFiT & \raggedright A text representation model based on general-domain language model pre-train, target task language model and classifier fine-tune. & \citep{Howard2018U}&$\surd$\\
 \midrule
\ac{LIR} & \raggedright A text representation model based on the local interaction representation. & This paper&$\times$\\
\ac{GIR} & \raggedright A text representation model based on the global interaction representation. & This paper&$\times$\\
\ac{HIR} & \raggedright A text representation model based on the hybrid interaction representation. & This paper&$\times$\\
 \midrule
LIR$_B$ & \raggedright A text representation model based on the local interaction representation model in the BERT fine-tuning architecture.
 & This paper&$\surd$ \\
 GIR$_B$ & \raggedright A text representation model based on the global interaction representation model in the BERT fine-tuning architecture.
 & This paper &$\surd$\\
HIR$_B$ & \raggedright A text representation model based on the hybrid interaction representation model in the BERT fine-tuning architecture.
 & This paper &$\surd$\\
LIR$_P$ & \raggedright A text representation model based on the local interaction representation model in the Pre-train Interact Fine-tune architecture.
& This paper &$\surd$\\
GIR$_P$ & \raggedright A text representation model based on the global interaction representation model in the Pre-train Interact Fine-tune architecture.
& This paper &$\surd$\\
HIR$_P$ & \raggedright A text representation model based on the hybrid interaction representation model in the Pre-train Interact Fine-tune architecture.
& This paper &$\surd$\\
\bottomrule
    \end{tabular}
  \label{models}
\end{footnotesize}
\end{table*}

\begin{table*}[h]
\centering
\caption{Dataset statistics.}
\label{table_s}
\begin{tabular}{l@{}rrrrr@{}}
\toprule
Dataset & IMDb & Yelp& TREC & AG & DBpedia\\
 \midrule
 type & sentiment & sentiment  & question  & topic & topic  \\
 \# training documents & 25K  & 560K & 5\phantom{.0}K   &  120\phantom{.6}K & 560K \\
 \# text documents     & 2K   & 50K  & 0.5K &  \phantom{12}7.6K & 70K \\
 \# classes            & 2\phantom{K}     &  5\phantom{K}    &  6\phantom{.5K}    &   4\phantom{K}    &  14\phantom{K}  \\
  \bottomrule
\end{tabular}
\end{table*}

\subsection{Model summary and research questions}
\label{model summary}

Table \ref{models} list the models to be discussed.
Among these models, LSTM, Char-level CNN, LIR, GIR and HIR models are neural based representation and don't experience the pretrain-finetuning process. 

\begin{enumerate}[align=left, leftmargin=*]
\item [\textbf{Baselines}]Four state-of-the-art baselines: two neural based representation model (i.e., LSTM~\citep{Liu2016Recurrent}, C-CNN~\citep{Zhang2015Character}), two language model pre-training based representation model (i.e., CoVe~\citep{McCann2017L}, ULMFiT~\citep{Howard2018U}).
\item [\textbf{Our proposals}] Nine flavors of approaches that we introduce in this paper: three interaction representation models (i.e., \ac{LIR}, \ac{GIR} and \ac{HIR}), three interaction representation models in the BERT architecture (i.e.,  LIR$_B$, GIR$_B$, HIR$_B$)
and the \acf{PIF} architecture (i.e., LIR$_P$, GIR$_P$, HIR$_P$).
\end{enumerate}

To assess the quality of our proposed interaction representation models and the \ac{PIF} architecture, we consider a text classification task and seek to answer the following questions:

\begin{enumerate}[align=left, leftmargin=*]
\item[\bf{RQ1}] Does the interaction representation incorporated in the text representation model help to improve the performance for text classification?
\item[\bf{RQ2}] Compared with the existing pretrain-finetuning approaches, does our proposed \ac{PIF} architecture help to improve the model performance for text classification?
\item[\bf{RQ3}] How does the trade-off parameter between \ac{LIR} and \ac{GIR} (as encoded in $\gamma$) impact the performance of  \ac{HIR} related model in terms of classification accuracy?
\item [\bf{RQ4}] Is the performance of our proposal sensitive to the length $L$ of text to be classified?
\end{enumerate}

\subsection{Datasets}
\label{datasets}
We evaluate our proposal on five publicly available datasets used in different application domains, e.g., sentiment analysis, questions classification and topic classification, which are widely used by the state-of-the-art models for text classification, e.g., {CoVe} \citep{McCann2017L} and  {ULMFiT} \citep{Howard2018U}.
Table \ref{table_s} details the statistics of the datasets.
We use accuracy as the evaluation metric to compare the performance of discussed models.

\begin{enumerate}[align=left]
\item[\textbf{Sentiment analysis}] Sentiment analysis mainly concentrates on the movie review and shopping review datasets.
For example, IMDb dataset proposed by \citep{Maas2011L} is a movie review dataset with binary sentimental labels.
While Yelp dataset compiled by \citep{Zhang2015Character} is a shopping review dataset that has two versions, i.e., binary and five-class version. 
We concentrate on the five-class version \citep{Johnson2017D}.
\item[\textbf{Question classification}] For question classification, \citet{Voorhees1999T} collected open-domain fact-based questions and divided them into broad semantic categories, which has six-class and fifty-class versions.
We mainly focus on the small six-class version and hold out 452 examples for validation and leave 5,000 for training, which is similar to \citep{McCann2017L}.

\item[\textbf{Topic classification}] For topic classification, we evaluate our proposals on the task of news article and ontology classification.
We use the AG news corpus collected by \citet{Zhang2015Character}, which has four classes of news with only the titles and description fields.
In addition, the DBpedia dataset,   collected by \citet{Zhang2015Character},  is used, which contains the title and abstract of each Wikipedia article with 14 non-overlapping ontology classes.
In general, the dataset division is the same as  in~\citep{Zhang2015Character}.
\end{enumerate}

\subsection{Model configuration and training}
\label{model configuration}
For data preprocessing, we split the text into sentences and tokenized each sentence using Stanford's CoreNLP \citep{Manning2014T}.
In addition, we discard the words with single characters and other punctuation and convert the upper-case letters ton the lower-cases letters.
In order to fit in the BERT pre-training, we add a special token for each sentence, e.g., [CLS] and [SEP].
The other data preprocessing follow the same way as \citep{Johnson2017D}

For model configuration, we use the same set of hyper-parameters across all datasets to evaluate the robustness of our proposal.
In the process of pre-training, we directly employ the trained $BERT_{base}$\footnote{https://github.com/google-research/bert} as our language model pre-training layer for simplicity.
As for the feature-based process, we follow the ELMo model\footnote{https://github.com/allenai/bilm-tf} and employ  AWD-LSTM \citep{Merity2018R} on the trained BERT layer to get the context-aware word embeddings.
For classifier fine-tuning layer, we adopt a softmax classifier and set the size of hidden layer to 100.
In addition, we set the dimension of word embeddings and hidden representation in the interaction representation layer to 400 and 200, respectively. 
We also apply a dropout of 0.4 to layers and 0.05 to the embedding layers.

For the whole training process, we use a batch size of 64, a base learning rate of 0.004 and 0.01 for fine-tuning the interaction representation layer and the classifier fine-tuning layer, respectively.
We employ a batch normalization mechanism \citep{Ioffe2015B} to accelerate the training of the neural networks.
Gradient clipping is applied by scaling gradients when the norm may exceed a threshold of 5 \citep{Pascanu2013O}.
For the fine-tuning process, we adopt the gradual unfreezing strategy \citep{Howard2018U} to fine-tune all neural layers.

\section{Results and Discussion}
In Section~\ref{PC}, we examine the performance of our proposal incorporated with the interaction representation and the \ac{HLMPf} on five public datasets, which aims at answering \textbf{RQ1} and \textbf{RQ2}.
Then, in Section~\ref{pa}, we analyze the impact of the trade-off parameter $\gamma$ in \ac{HIR} related model to answer \textbf{RQ3}.
Finally, to answer \textbf{RQ4}, section Section~\ref{dis:part3} focuses on investigating the impact on the text classification by varying the text length.

\subsection{Performance comparison}
\label{PC}

\subsubsection{Performances about the interaction representation}
\label{PATIR}

To answer \textbf{RQ1}, we first compare the performance of the basic interaction representation based models (i.e., LIR, GIR and HIR) with the baselines and present the results in Table \ref{table_inter}

\begin{table}[h]
\centering
\caption{Error rate (\%) about the interaction representation on different datasets. 
(The results of the best baseline and the best performer in each column are underlined and boldfaced, respectively.
Results marked with $*$ are re-printed from \citep{Zhang2015Character,McCann2017L,Howard2018U,Zhou2016T}. The rest are obtained by our own implementation.
{Statistical significance of pairwise differences (the best proposed model vs.\ the best neural-network-based baseline) are determined by a $t$-test ($^\blacktriangle$/$^\blacktriangledown$ for $\alpha$ = .01)}}
\label{table_inter}
\begin{tabular}{llllll}
\toprule
Datasets & IMDb & Yelp & TREC & AG & DBpedia   \\
\midrule
\ac{LSTM}& 8.72 & $41.83^*$ & 7.66 & $13.94^*$ & $1.45^*$\\
C-CNN&  7.36 & $37.95^*$ & 6.48 & $9.51^*$ & $1.55^*$\\
CoVe & $8.2^*$ & -- & $4.2^*$ & -- & -- \\
ULMFiT & $\bm {\underline{4.6}^*}$ & $\bm{\underline{29.98}^*}$ & $\bm{\underline{3.6}^*}$ & $\bm{\underline{5.01}^*}$ & $\bm{\underline{0.80}^*}$ \\
\midrule
\ac{LIR}& 6.86   &  35.58  &  5.76  & 7.83 & 1.31  \\
\ac{GIR} & 6.92 & 35.46 & 5.87 & 8.20 & 1.37 \\
\ac{HIR}& {6.73}$^\blacktriangle$ & {34.18}$^\blacktriangle$ & {5.44}$^\blacktriangle$ & {7.53}$^\blacktriangle$ & {1.24}$^\blacktriangle$ \\

\bottomrule
\end{tabular}
\end{table}

As to the baselines, we present two types of representation models, i.e., the neural-network based model (LSTM and C-CNN) and the pretrain-finetuning based model (CoVe and ULMFiT).
For the neural-network based model, C-CNN achieves a better performance than LSTM.
While in the pretrain-finetuning based model, ULMFiT is obviously the better one.
Interestingly, comparing these two types of models, we can find that the representation models with pretrain-finetuning process have super advantages in terms of reducing error rate.
Specially,  with regard to C-CNN, ULMFiT reduces the error dramatically by 37.5\%, 26.6\%, 44.4\%, 89.8\% and 48.4\% on the corresponding datasets (IMDb, Yelp, TREC, AG and DBpedia in order, which is the same in the following text).
This may be due to the fact that  the pre-training on a large text corpora can capture the deep syntactic and semantic information, which cannot be realized by only training on the neural networks.   

Similarly, our proposals only with the interaction representation, i.e., \ac{LIR}, \ac{GIR} and \ac{HIR},  cannot beat the state-of-the-art pretrain-finetuning based model, i.e., ULMFiT.
But for the neural-network based baselines, our proposals can achieve better performance in terms of {error rate}.
In particular, \ac{HIR} is the best performing model among our proposals, which shows an improvement against the best neural-network based baseline, i.e.,  C-CNN, resulting in 8.6\%, 9.9\%, 16.0\%, 26.3\% and 20\% reduction in terms of error rate on the respective datasets.
LIR and GIR, following HIR, can outperform C-CNN on all datasets.
The aforementioned findings indicate that compared with the traditional neural-network based models, modeling the interaction process explicitly can better capture the semantics relation between source elements in the text and generate more meaningful text representation.
Especially for HIR, by representing the local and global interaction between words, it is more effective to improve the performance of the downstream applications.

\subsubsection{Performances about the pretrain-finetuning}
\label{PATRP}

In section \ref{PATIR}, the effectiveness of the pretrain-finetuning based and the interaction-related models have been proven.
However, the basic interaction representation based models cannot beat the state-of-the-art pretrain-finetuning based model, i.e., ULMFiT.
Hence, we incorporate them with the popular pretrai-finetuning architecture (i.e., BERT) and our \ac{PIF} architecture to get the corresponding models (i.e., LIR$_B$, GIR$_B$, HIR$_B$ and LIR$_P$, GIR$_P$, HIR$_P$), respectively.
To answer \textbf{RQ2}, we compare the performance of these proposed models with ULMFiT and present their experimental results in Table \ref{table_pretrain}.

\begin{table}[h]
\centering
\caption{Error rate (\%) about the pretrain-finetuning process on different datasets. (The results of the best baseline and the best performer in each column are underlined and boldfaced, respectively.
Results marked with $*$ are re-printed from \citep{Howard2018U}. The rest are obtained by our own implementation.
{Statistical significance of pairwise differences (the best proposed model vs.\ the best neural-network-based baseline) are determined by a $t$-test ($^\blacktriangle$/$^\blacktriangledown$ for $\alpha$ = .01)}}

\label{table_pretrain}
\begin{tabular}{llllll}
\toprule
Datasets & IMDb & Yelp & TREC & AG & DBpedia   \\
\midrule
ULMFiT & $\underline{4.6}^*$ & $\underline{29.98}^*$ & $\underline{3.6}^*$ & $\underline{5.01}^*$ & $\underline{0.80}^*$ \\
LIR$_B$  & 4.58 & 28.42 & 3.54 & 4.93 & 0.81 \\
GIR$_B$  & 4.69 & 28.84 & 3.55 & 5.03 & 0.84\\
HIR$_B$  & 4.24 & 28.31 & 3.37 & 4.88 &  0.78 \\
\midrule
LIR$_P$  & 4.25 & 27.33 & 3.40  & 4.92 & 0.80 \\
GIR$_P$  & 4.31 & 27.66 & 3.48  & 4.94 & 0.81 \\
HIR$_P$  & \textbf{4.04}$^\blacktriangle$ & \textbf{27.07}$^\blacktriangle$ & \textbf{3.33}$^\blacktriangle$ & \textbf{4.85}$^\blacktriangle$ &  \textbf{0.77}$^\blacktriangle$ \\

\bottomrule
\end{tabular}
\end{table}

Clearly, as shown in Table \ref{table_pretrain}, our basic interaction-related models incorporated with the pretrain-finetuning process generally outperform the state-of-the-art model, i.e., ULMFiT, except for some  cases, e.g., the LIR$_B$ on DBpedia,  GIR$_B$ on AG and DBpedia, GIR$_P$ on DBpedia.
%
This findings again prove that our basic interaction representation models have the promising perspectives under the pretrain-finetuning architecture. 
With regard to the BERT architecture, our interaction-related models present the similar accuracy distribution to the basic interaction representation models in Table \ref{table_inter}.
HIR$_B$ is the best performer using the BERT architecture, followed by LIR$_B$ and GIR$_B$.
Specially, for each dataset, HIR$_B$ shows an obvious improvement of 7.9\%, 5.6\%, 6.4\%, 2.6\% and 2.5\% against ULMFiT, respectively.
While LIR$_B$, except on the DBpedia, also gains a minor improvement of 0.4\%, 5.2\%, 1.7\%, 1.6\% against ULMFiT, respectively.
GIR$_B$, a bit worser than LIR$_B$, beats the ULMFiT on 3 out of 5 datasets.

The similar findings can also be found in the PIF architecture.
In particular, HIR$_P$ achieves the best performance not only in the PIF architecture but among all discussed models.
Compared with the baseline ULMFiT,  HIR$_P$ gains substantial improvements of 12.1\%, 9.7\%, 7.5\%, 3.2\%, 3.8\% in terms of {error rate} on respective datasets.
{In addition, LIR$_P$ wins the comparisons against ULMFiT, resulting in 7.6\%, 8.8\%, 5.6\%, 1.8\%  improvements on the respective datasets and an equal performance on DBpedia.}
While GIR$_P$ defeats the ULMFiT model on 4 out 5 datasets.

Furthermore, comparing the same type of interaction models with different architectures (e.g., type LIR: LIR, LIR$_B$, LIR$_P$), we can find that there exists a unchanged ranking order of performance on each dataset, i.e., LIR$_P$ $>$ LIR$_B$ $>$ LIR, GIR$_P$ $>$ GIR$_B$ $>$ GIR, HIR$_P$ $>$ HIR$_B$ $>$ HIR.
This ranking order demonstrates that our proposed PIF architecture that combines the feature-based and fine-tuning based strategies is the most effective architecture, followed by the fine-tuning based strategy, BERT.
While the neural-network based models are worse than the former kinds of models.

\subsection{Parameters analysis}
\label{pa}

\begin{figure}[!t]
        \centering
        \begin{subfigure}[t]{0.6\textwidth}
         \centering
                \includegraphics[clip,trim=0mm 2mm 0mm 0mm,width=1.0\textwidth]{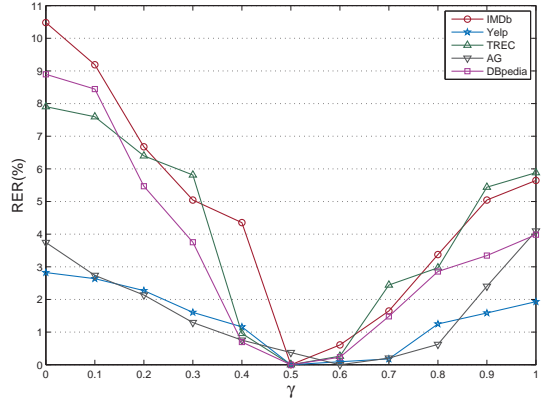}
                \caption{HIR performance on each dataset.}
                \label{HIR_G}
        \end{subfigure}
        \begin{subfigure}[t]{0.6\textwidth}
         \centering
                \includegraphics[clip,trim=0mm 2mm 0mm 0mm,width=1.0\textwidth]{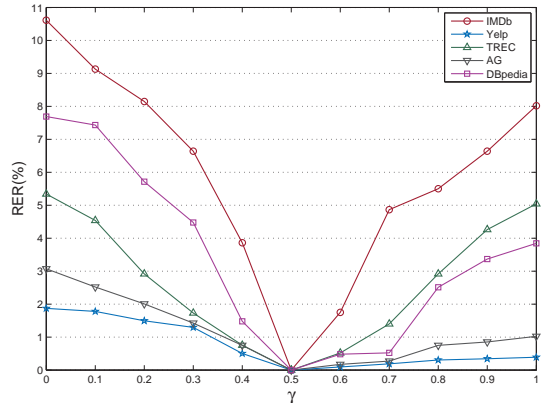}
                \caption{HIR$_B$ performance on each dataset.}
                \label{HIRB_G}
        \end{subfigure}
                \begin{subfigure}[t]{0.6\textwidth}
                 \centering
                        \includegraphics[clip,trim=0mm 2mm 0mm 0mm,width=1.0\textwidth]{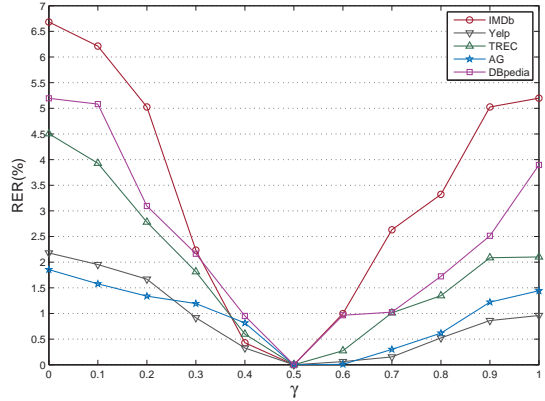}
                        \caption{HIR$_P$ performance on each dataset.}
                        \label{HIRP_G}
                \end{subfigure}
        \caption{Effect on performance of the \ac{HIR} related models in terms of
        \ac{RER} by changing the trade-off parameter $\gamma$, tested on all datasets.}
\label{G_change}
\end{figure}

Next we turn to \textbf{RQ3} and conduct a parameter sensitivity analysis of our HIR related models, i.e., HIR, HIR$_B$ and HIR$_P$.
{Clearly, as shown in Table \ref{table_inter} and Table \ref{table_pretrain}, for different datasets, the same model has varied error rates on different orders of magnitude, e.g., HIR on IMDb and Yelp ($6.73 ~vs~ 34.18$).
To better present the $\gamma$  effect of the same model on different datasets, we introduce an evaluation metric, \ac{RER}, which is defined as, given a dataset, the relative improvement ratio of the lowest error rate with regard to the others with different $\gamma s$.}
In addition, we examine the performances of these three models in terms of \ac{RER} by gradually changing the parameters $\gamma$ from 0 to 1 with an interval 0.1.
We plot the \ac{RER} results of HIR, HIR$_B$ and HIR$_P$ in Figure \ref{HIR_G}, Figure \ref{HIRB_G} and Figure \ref{HIRP_G}, respectively.

As shown in Figure \ref{HIR_G}, HIR achieves the lowest error rate when $\gamma = 0.5$ on all datasets (except $\gamma =0.6$ for Yelp dataset), which is $0$ in the figure.
In addition, the \ac{RER} of HIR on each dataset decreases consistently when $\gamma$ varies from $0$ to $0.5$ ($0.6$ for Yelp);
after that, the \ac{RER} metric goes up when $\gamma$ changes from $0.5$ ($0.6$ for Yelp) to $1$.
The similar phenomena can be found in Figure \ref{HIRB_G} and Figure \ref{HIRP_G}.
HIR$_B$ and HIR$_P$ both achieve the lowest error rate when $\gamma =0.5$.
In addition,  the \ac{RER} of these two models on each dataset first keeps a stable decrease to the lowest point $0$ and then increases stably until $\gamma =1$.  

Interestingly, comparing the curve gradient on both sides of $\gamma=0.5$, we can find that the gradient of the left side is steeper than that of the right side, which indicates that GIR can result in the increase of error rate more easily than {LIR}.
{Furthermore, comparing the same model on different datasets, we can find that the change ranges of RER on IMDb, DBpedia and TREC,  is greater than that on Yelp and AG.
The phenomena may be due to the differences of statistical characteristics among these datasets, which require further experiments to find potential reasons.}

\begin{figure}[!t]
  \centering
   \includegraphics[width=0.6\textwidth]{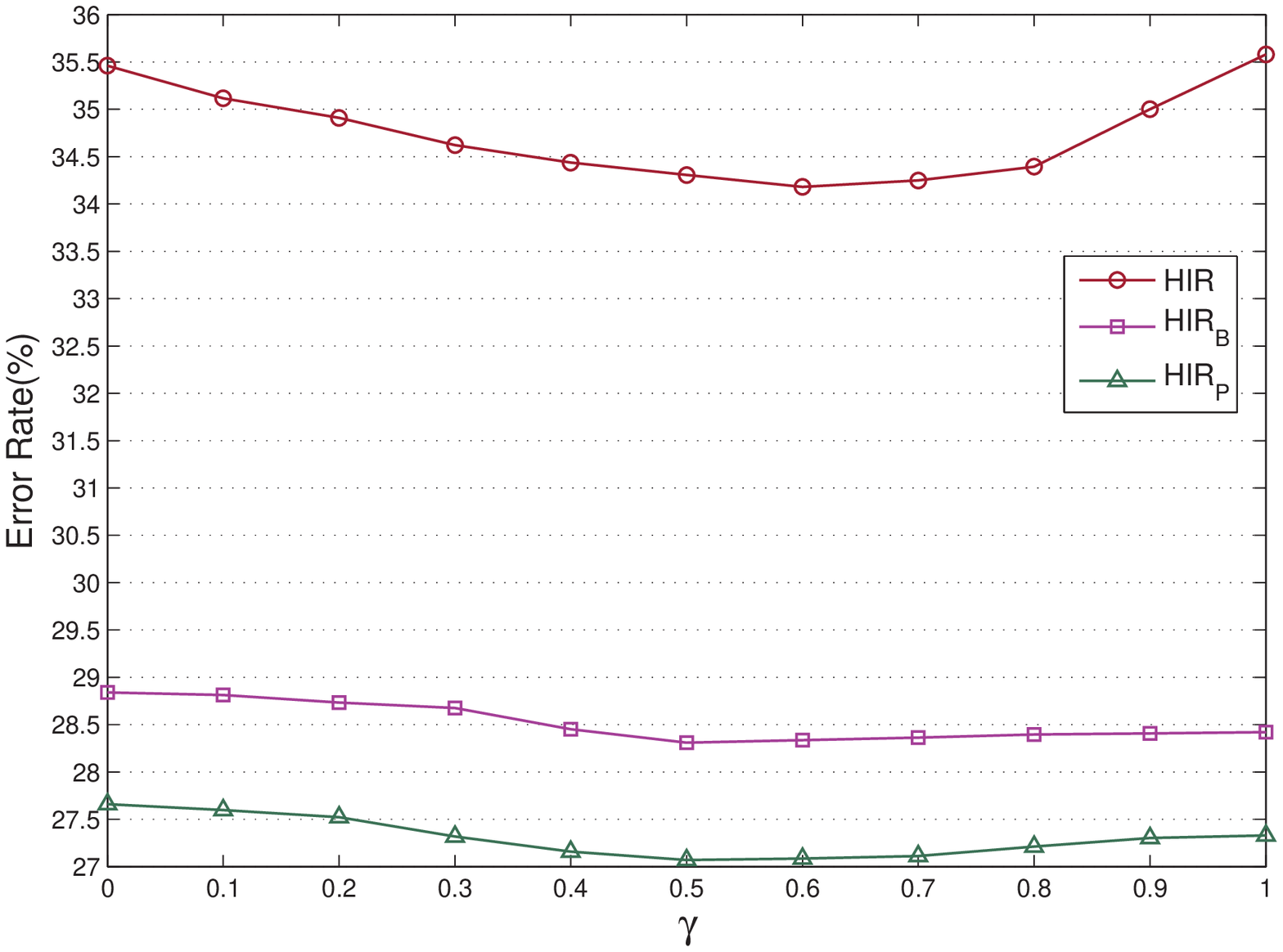}
   \caption{Effect on the performance of \ac{HIR} related models in terms of error rate by changing the trade-off parameter $\gamma$, tested on the Yelp dataset.}
\label{Yelp_G}
\end{figure}

Curiously, we also want to find whether the relation HIR$_P$ > HIR$_B$ > HIR can always keep unchanged when the trade-off parameter $\gamma$ increases from $0$ to $1$.
Due to the text space, we only select the dataset Yelp as the analytical object, which has the highest error rate among these datasets.
We plot the experimental results in Figure \ref{Yelp_G}.
Clearly, as Figure \ref{Yelp_G} shows, {we can find that the performance of HIR$_P$ is the lowest in terms of error rate, followed by HIR$_B$, and the highest is HIR, when $\gamma$ increases from $0$ to $1$.}
This result is consistent with the previous finding HIR$_P$ $>$ HIR$_B$ $>$ HIR, i.e., the effectiveness of our \ac{PIF} architecture.
On the other hand, it indicates that the effectiveness of our PIF architecture is not sensitive to the trade-off parameter $\gamma$.

\subsection{Impact of the text length}
\label{dis:part3}
\begin{figure}[t]
  \centering
   \includegraphics[width=0.6\textwidth]{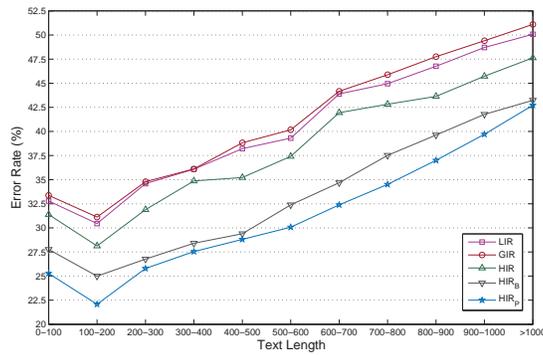}
   \caption{Effect on the performance of interaction related models in terms of error rate with varied text length, tested on Yelp dataset. }
\label{yelp length}
\end{figure}

To answer \textbf{RQ4}, we manually group the text according to the text length $L$, e.g., 0--100, 100--200, $\dots$, 900--1000, $>$1000.
We campare the performance of interaction representation related models, e.g., LIR, GIR, HIR, HIR$_B$ and HIR$_P$, under different settings of 
text length.
We plot this experimental results in Figure \ref{yelp length}

Clearly, as shown in Figure \ref{yelp length}, we can find the relation LIR $>$ GIR $>$ HIR $>$ HIR$_B$ $>$ HIR$_P$ unchanged when text length increases.
This phenomenon is consistent with the findings in Section~\ref{PATIR}, which indicates the effectiveness of interaction representation and PIF architecture is not affected by the text length.

Interestingly, as the text length increases, the performances of all discussed models decrease first to reach the lowest error rate at the point of group 100--200, and then keep a constant increase.
This finding may be explained by the fact that the longer the text, the richer the information it provides, which results in targeting the class label of text more easily, i.e., the decrease of error rate in the earlier stage.
But as the text length grows, the structure and semantics of text become more complex and variable, the proposed models find it harder to  get the exact representation.

\section{Conclusion and Future Work}
In this paper, we focus on the task of text classification and propose a novel pipeline, the PIF architecture, which incorporates the respective advantages from feature based and fine-tuning based strategies in the language model pretrain-finetuning process.
We also introduce the concept of interaction representation and propose a two-perspective interaction representation for sentence embeddings, i.e., a local interaction representation (LIR) and a global interaction representation (GIR).
We combine these two representations to produce a hybrid interaction representation model, i.e., HIR.

We evaluate these models on five widely-used datasets for text classification. 
Our experimental results shows that:
\begin{inparaenum}[(1)]
(1) compared with the traditional neural-network based models, our basic interaction-related models can help boost the performance for text classification  in terms of error rate.
(2) our proposed PIF architecture is more effective to help improve the text classification than the existing feature-based as well as the fine-tuning based strategies.
Specially, HIR$_P$ model present the best  performance on each dataset.
(3) the effectiveness of interaction representation and the PIF architecture is not affected by the text length.
\end{inparaenum}

As to future work, we plan to evaluate our models for other tasks so as to verify the robustness of the interaction representation models.
In addition, the existing fine-tuning approach is too general. 
We want to investigate some task-sensitive fine-tuning methods to better improve the performance.

\section{Acknowledgements}
This work was partially supported by
the National Natural Science Foundation of China under No.\ 61702526,
the Defense Industrial Technology Development Program under No.\ JCKY2017204B064,
the National Advanced Research Project under No.\ 6141B0801010b,
Ahold Delhaize,
the Association of Universities in the Netherlands (VSNU),
and
the Innovation Center for Artificial Intelligence (ICAI).
All content represents the opinion of the authors, which is not necessarily shared or endorsed by their respective employers and/or sponsors.

\bibliography{Interaction.bib}

\end{document}